\newcommand{\be}{\begin{equation}}
\newcommand{\ee}{\end{equation}}
\newcommand{\bea}{\begin{eqnarray}}
\newcommand{\eea}{\end{eqnarray}}

\documentclass[10pt,aps,prb,twocolumn,sort&compress,superscriptaddress,floatfix]{revtex4-1}
\usepackage{microtype} 
\usepackage{graphicx} 
\usepackage{dcolumn} 
\usepackage{natbib}
\usepackage{units}
\usepackage{amsmath}
\usepackage[colorlinks=true,citecolor=blue,linkcolor=blue,pdfborder={0 0 0}]{hyperref}

\begin{document}

\title{Spin pumping damping and magnetic proximity effect in Pd and Pt spin-sink layers}

\author{M. Caminale}\email{michael.caminale@cea.fr}
\affiliation{Fondation Nanosciences, F-38000 Grenoble, France}
\affiliation{SPINTEC, Univ. Grenoble Alpes / CEA / CNRS, F-38000 Grenoble, France}

\author{A. Ghosh}\thanks{Present address: Data Storage Institute, Agency for Science, Technology and Research (A*STAR), Singapore 138634}
\author{S. Auffret}
\author{U. Ebels}
\affiliation{SPINTEC, Univ. Grenoble Alpes / CEA / CNRS, F-38000 Grenoble, France}

\author{K. Ollefs}
\affiliation{Fakult\"at f\"ur Physik and Center for Nanointegration (CENIDE), Universit\"at Duisburg-Essen, 47057 Duisburg, Germany}

\author{F. Wilhelm}
\author{A. Rogalev}
\affiliation{European Synchrotron Radiation Facility (ESRF), 38054 Grenoble Cedex, France}

\author{W.E. Bailey}\email{web54@columbia.edu}
\affiliation{Fondation Nanosciences, F-38000 Grenoble, France}
\affiliation{Dept. of Applied Physics \& Applied Mathematics, Columbia University, New York NY 10027, USA}

\date{\today}

\begin{abstract}
We investigated the spin pumping damping contributed by paramagnetic layers (Pd, Pt) in both direct and indirect contact with ferromagnetic Ni$_{81}$Fe$_{19}$ films.
We find a nearly linear dependence of the interface-related Gilbert damping enhancement $\Delta\alpha$ on the heavy-metal spin-sink layer thicknesses t$_\textrm{N}$ in direct-contact Ni$_{81}$Fe$_{19}$/(Pd, Pt) junctions, whereas an exponential dependence is observed when Ni$_{81}$Fe$_{19}$ and (Pd,  Pt) are separated by \unit[3]{nm} Cu. 
We attribute the quasi-linear thickness dependence to the presence of induced moments in Pt, Pd near the interface with Ni$_{81}$Fe$_{19}$, quantified using X-ray magnetic circular dichroism (XMCD) measurements.
Our results show that the scattering of pure spin current is configuration-dependent in these systems and cannot be described by a single characteristic length.
\end{abstract}

\maketitle

\section{Introduction}

As a novel means of conversion between charge- and spin- currents, spin Hall phenomena have recently opened up new possibilities in magneto-electronics, with potential applications in mesocale spin torques and electrical manipulation of domain walls \citep{Weber2001,Stiles2002,Saitoh2006,Miron2011,Liu2011a,Feng2012,Nakayama2012,Ryu2013,Rojas-Sanchez2014}. 
However, several aspects of the scattering mechanisms involved in spin current flow across thin films and interfaces are not entirely understood. 
Fundamental studies of spin current flow in ferromagnet/non-magnetic-meal (F/N) heterostructures in the form of continuous films have attempted to isolate the contributions of interface roughness, microstructure and impurities \citep{Zhang2011,Zhang2015,Tokac2015}. 
magnet/non-magnetic-meal (F/N) heterostructures in the form of continuous films have attempted to isolate the contributions of interface roughness, microstructure and impurities \citep{Zhang2011,Zhang2015,Tokac2015}. 
Prototypical systems in this class of studies are Ni$_{81}$Fe$_{19}$/Pt (Py/Pt) \citep{Saitoh2006,Liu2011a,Ghosh2011,Ghosh2012a,Nakayama2012,Feng2012,Niimi2013,Boone2015,Nan2015,Ruiz-Calaforra2015} and Ni$_{81}$Fe$_{19}$/Pd (Py/Pd) \citep{Tserkovnyak2002,Foros2005,Ghosh2012a,Ryu2013,Zhang2015,Boone2015} bilayers.
In these systems, Pt and Pd are employed either as efficient spin-sinks or spin/charge current transformers, in spin pumping and spin Hall experiments, respectively. 
Pd and Pt are metals with high paramagnetic susceptibility and when placed in contact with a ferromagnetic layer (eg. Py, Ni, Co or Fe) a finite magnetic moment is induced at the interface by direct exchange coupling \citep{Lee2007,Bailey2012,Wilhelm2000,Vogel97}.

The role of the magnetic proximity effect on interface spin transport properties is still under debate.
Zhang \emph{et al.} \citep{Zhang2015a} have reported that induced magnetic moments in Pt and Pd films in direct contact with Py correlate strongly reduced spin Hall conductivities. 
This is ascribed to a spin splitting of the chemical potential and on the energy dependence of the intrinsic spin Hall effect.
In standard spin pumping theory \citep{Tserkovnyak2005}, possible induced moments in N are supposed to be \emph{a priori} included in calculations of the spin-mixing conductance $g^{\uparrow\downarrow}$ of a F\textbar N interface \citep{Brataas2000,Zwierzycki2005}, which tends to be insensitive to their presence. 

Recent theoretical works, on the other hand, propose the need of a generalized spin pumping formalisms including spin flip and spin orbit interaction at the F\textbar N interface, in order to justify discrepancies between experimental and calculated values of mixing conductance \citep{Liu2014,Chen2015a}. 
At present, it is still an open issue whether and how proximity-induced magnetic moments in F/N junctions are linked to the variety of the spin-transport phenomena reported in literature \citep{Zhang2011,Sun2013,Nan2015}. 

Here, we present an experimental study of the prototypical systems: Py/Pd, Pt and Py/Cu/Pd, Pt heterostructures.
The objective of our study is to address the role of proximity induced magnetic moments in spin pumping damping. 
To this end, we employed two complementary experimental techniques.
X-ray magnetic circular dichroism (XMCD) is an element sensitive technique which allows us to quantify any static proximity-induced magnetic moments in Pt and Pd.
Ferromagnetic resonance (FMR) measurements provide indirect information on the spin currents pumped out the Py layer by the precessing magnetization, through the characterization of the Pd, Pt thickness dependence of the interface-related Gilbert damping $\alpha$. 
In Fig. \ref{fig2} (Sec. \ref{FMR}), comparative measurements in Py/Cu/N and Py/N structures show a change of the N thickness dependence of $\Delta\alpha(t_\textrm{N})$ from an exponential to a linear-like behavior.
A change in $\Delta\alpha(t_\textrm{N})$ indicates a transformation in the spin scattering mechanism occurring at the interface, ascribed here to the presence of induced moments in directly exchange coupled F/N systems.
Theoretical works predicted a deviation from a conventional N-thickness dependence when interface spin-flip scattering is considered in the pumping model \citep{Liu2014,Chen2015a}, however no functional form was provided. 
For Py/N systems, we find that the experimental thickness dependence cannot be described by standard models \citep{Tserkovnyak2005,Boone2013,Boone2015}, but rather a linear function reproduces the data to a better degree of accuracy, by introducing a different characteristic length.
We speculate that the spatial extent of spin current absorption in F/N systems shows an inverse proportionality to interfacial exchange coupling energy, obtained from XMCD, as proposed before for spin polarized, decoupled interfaces in  F$_1$/Cu/F$_2$ heterostructures \citep{Ghosh2012a}.

\section{Experiment}

The heterostructures were fabricated by DC magnetron sputtering on ion-cleaned Si/SiO$_{2}$ substrates in the form of \emph{substrate/seed/multilayer/cap} stacks, where Ta(\unit[5]{nm})/Cu(\unit[5]{nm}) bilayer was employed as \emph{seed}. 
Ta/Cu is employed to promote $<111>$ growth in Py and subsequent fcc layers (Pd, Pt), and Ta is known to not affect the damping strongly \citep{Boone2013,Liu2012a,Nan2015}.
Different stacks were grown as \emph{multilayer} for each measurement.

For FMR measurements, we have \emph{multilayer} = Py($t_\textrm{F}$)/N($t_\textrm{N}$), Py($t_\textrm{F}$)/Cu(\unit[3]{nm})/N($t_\textrm{N}$) with N = Pd, Pt; an Al(\unit[3]{nm}) film, oxidized in air, was used as \emph{cap}.
The smallest N layer thickness $t_\textrm{N}$ deposited is \unit[0.4]{nm}, the maximum interdiffusion length observed for similar multilayers \citep{Kohlhepp2002}.
Samples with \emph{multilayer} = Py($t_\textrm{N}$)/Cu(\unit[3]{nm}) and no sink layer were also fabricated as reference for evaluation of the Gilbert damping enhancement due to the Pd or Pt layer.
The $t_\textrm{N}$-dependence measurements of FMR were taken for Py thicknesses $t_\textrm{F}=$ \unit[5] and \unit[10]{nm}.
Results from the $t_\textrm{F}=$ \unit[10]{nm} data set are shown in Appendix \ref{damping_section}.
Measurements of the FMR were carried out at fixed frequency $\omega$ in the 4-24 Ghz range, by means of an in-house apparatus featuring an external magnetic field up to \unit[0.9]{T} parallel to a coplanar waveguide with a broad center conductor width of \unit[350]{$\mu$m}.

For XMCD measurements, given the low X-ray absorption cross-section presented by Pt and Pd absorption edges, a special set of samples was prepared, consisting of 20 repeats per structure in order to obtain sufficiently high signal-to-noise ratio. 
In this case, we have \emph{multilayer} = [Py(\unit[5]{nm})/N]$_{20}$, with N = Pd(\unit[2.5]{nm}) and Pt(\unit[1]{nm}); Cu(\unit[5]{nm})/Py(\unit[5]{nm})/Al(\unit[3]{nm}) was deposited as \emph{cap}. 
The Pt and Pd thicknesses were chosen to yield a damping enhancement equal about to half of the respective saturation value (as it will be shown later), i.e. a thicknesses for which the F/N interface is formed but the damping enhancement is still increasing. 
XMCD experiments were carried out at the Circular Polarization Beamline ID-12 of the European Synchrotron Radiation Facility (ESRF) \citep{Rogalev15}. 
Measurements were taken in total fluorescence yield detection mode, at grazing incidence of 10$^{\circ}$, with either left or right circular helicity of the photon beam, switching a \unit[0.9]{T} static magnetic field at each photon energy value (further details on the method are in Ref. \citep{Bailey2012}).
No correction for self-absorption effects is needed; however XMCD spectra measured at the L$_{2,3}$ edges of Pd have to be corrected for incomplete circular polarization rate of monochromatic X-rays which is 12\% and 22\% at L$_{3}$ and L$_{2}$, respectively. 
The circular polarization rate is in excess of 95 \% at the L$_{2,3}$ edges of Pt.  

\section{Results and analysis}

In order to study how the proximity-induced magnetic moments may affect the absorption of spin-currents through interfaces, the static moment induced in Pt, Pd layers in direct contact with ferromagnetic Py in characterized first, by means of XMCD. 
The value of the induced moment extracted for the two Py/N systems is used to estimate the interfacial exchange energy acting on the two paramagnets.
Afterwards, the dynamic response of the magnetization is addressed by FMR measurements in Py/N (direct contact) and Py/Cu/N (indirect contact) heterostructures. 
From FMR measurements carried out on both configurations as a function of N thickness, the damping enhancement due to the presence of the spin-sink layers Pt and Pd is obtained from the frequency-dependence of the FMR linewidth.
The relation between the static induced moment and the spin pumping damping is discussed by comparing the results of the direct with indirect contact systems.

\begin{figure}
	\includegraphics[width=\columnwidth]{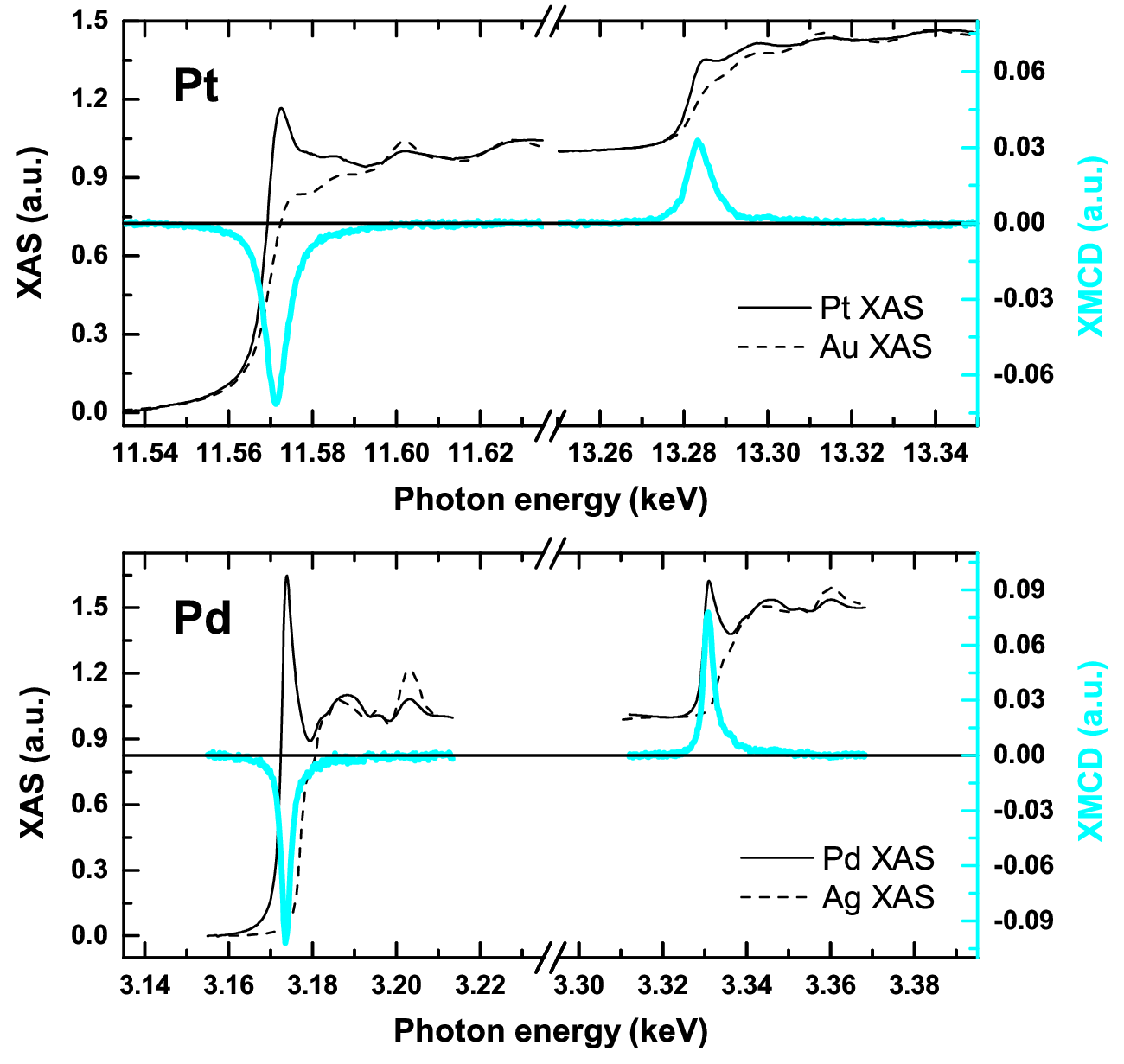}
	\caption{(Color online) X-ray absorption (XAS, left axis) and magnetic circular dichroism (XMCD, right axis) spectra at the L-edges of Pt (top panel) and Pd (bottom panel) for [Py(5nm)/Pt(\unit[1]{nm})]$_{20}$ and [Py(5nm)/Pd(\unit[2.5]{nm})]$_{20}$ multilayers. The dashed traces represent XAS spectra at L-edge of Ag and Au used as background of Pd and Pt, respectively, to extract the values of induced magnetic moment reported in Tab. \ref{table1}.}
	\label{fig1}
\end{figure}

\subsection{XMCD: Probing the induced magnetic moment}\label{XMCD}

In Fig. \ref{fig1} we report X-ray absorption (XAS) and magnetic circular dichroism (XMCD) spectra at the $L_{2,3}$ edges of Pt (top panel) and Pd (bottom panel) taken on Py(\unit[5]{nm})/Pt(\unit[1]{nm})$|_{20}$ and Py(\unit[5]{nm})/Pd(\unit[2.5]{nm})$|_{20}$, respectively. 
Rather intense XMCD signals have been detected at both Pt and Pd L$_{2,3}$ edges, showing unambiguously that a strong magnetic moment is induced by direct exchange coupling at the Py\textbar N interface. 
The static induced moment is expected to be ferromagnetically coupled with the magnetization in Py \citep{Lee2007}.
From the integrals of XMCD spectra, the induced magnetic moment on the Pt, Pd sites is determined by applying the sum rules as in Ref. \citep{Bailey2012} (and references therein). 
In Py/Pd(\unit[2.5]{nm})$|_{20}$, Pd atoms bear a moment of $\textrm{0.12}~\mu_B/\textrm{at}$, averaged over the whole volume of the volume, with an orbital-to-spin ratio $m_\textrm{L}/m_\textrm{S}=\textrm{0.05}$. 
In Py/Pt(\unit[1]{nm})$|_{20}$, a magnetic moment $\textrm{0.27}~\mu_B/\textrm{at}$ is found on Pt, comparable to that reported for Ni/Pt epitaxial multilayers \citep{Wilhelm2000}, with a relatively high orbital character $m_\textrm{L}/m_\textrm{S}=\textrm{0.18}$, as compared with Pd induced moment. 
The large difference in  volume-averaged induced moment per atom comes from the different film thickness, hence volume, for Pt and Pd.
Assuming that the induced magnetic moment is confined to the first atomic layers at the interface with Py \citep{Vogel97,Wilhelm2000}, one could estimate $\textrm{0.32}~\mu_B/\textrm{at}$ for Pd and $\textrm{0.30}~\mu_B/\textrm{at}$ for Pt \cite{Note1}.

\begin{figure}
	\includegraphics[width=0.8\columnwidth]{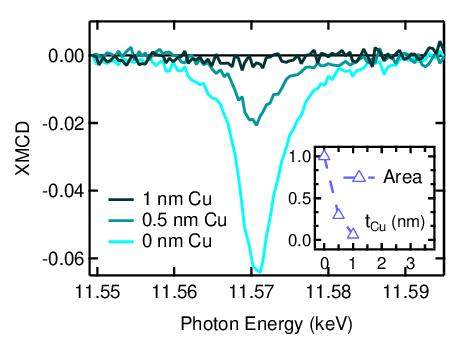}
	\caption{(Color online) XMCD spectra at the L$_{3}$ edge of Pt for [Py(5nm)/Cu(t$_{Cu}$)/Pt(\unit[1]{nm})]$_{15}$, with t$_{Cu}=0$, 0.5 and \unit[1]{nm}. As inset the ares of the peak is plot as a function of Cu thickness.}
	\label{fig_XMCD_Cu}
\end{figure}

When a \unit[3]{nm} thick Cu interlayer is introduced between Py and N, a two orders of magnitude smaller induced moment ($\textrm{0.0036}~\mu_B/\textrm{at}$) was found for \unit[2.5]{nm} Pd  \citep{Bailey2012}, while Pt showed an XMCD signal of the order of the experimental sensitivity, $\sim 0.5 \cdot 10^{-3}~\mu_B/\textrm{at}$. 
In Fig. \ref{fig_XMCD_Cu} XMCD spectra at the L$_{3}$ edge of Pt are shown for Cu interlayer thicknesses 0, 0.5 and \unit[1]{nm}. 
For \unit[0.5]{nm} Cu the integral of XMCD signal at the L$_{3}$ edge shrinks to 30$\%$, while for \unit[1]{nm} it is reduced to zero within experimental error. 
This result could be explained  either by a 3d growth of the Cu layer, allowing a fraction of the Pt layer to be in direct contact with Py for Cu coverages of \unit[0.5]{nm}, or by diffusion of magnetic Ni atoms in Cu on a scale shorter than \unit[1]{nm}. 
The film then becomes continuous, and at \unit[1]{nm} coverage, no direct exchange coupling takes place between Py and Pt layers. 
For FMR measurements presented in the following section, a \unit[3]{nm} thick Cu interlayer is employed, reducing also any possible indirect exchange coupling. 

\begin{table}
\begin{tabular}{c|c|c|c|c|c|c|c|c}
$N$& $\chi_\textrm{mol}$ \citep{l-b-paramag-4d5d}& S \citep{l-b-paramag-4d5d} & $N_0$ & a$_{bulk}$ & $t_\textrm{N}$ & $\langle M\rangle$ & $M_{i}$ &  $J_{ex}$ \\
& \footnotesize{[\nicefrac{\unit{cm$^{3}$}}{\unit{mol}}]} & & \footnotesize{[\nicefrac{1}{eV$\cdot$at}]} & \footnotesize{[\unit{nm}]} & \footnotesize{[\unit{nm}}] & \footnotesize{[\nicefrac{$\mu_B$}{at}]} & \footnotesize{[\nicefrac{$\mu_B$}{at}]}  & \footnotesize{[\unit{meV}]} \\
& \footnotesize{10$^{-4}$} & & & & & & \\ \hline 
Pd & 5.5\footnotesize{$\pm$0.2} 	& 9.3 & 0.83\footnotesize{$\pm$0.03} & 0.389 & 2.5 & 0.116	& 0.32 & 42 \\ 
Pt & 1.96\footnotesize{$\pm$0.1} & 3.7 & 0.74\footnotesize{$\pm$0.04} & 0.392 & 1.0 & 0.27 & 0.30 & 109 \\
\end{tabular}
\centering
\caption{Spin-sink layer N properties in Py/N heterostructures: experimental molar susceptibility $\chi_\textrm{mol}$ at 20 $^\circ$C; density of states $N_0$ calculated from tabulated $\chi_\textrm{mol}$; Stoner parameter S from Ref. \citep{l-b-paramag-4d5d}; bulk lattice parameter $a$; layer thicknesses $t_\textrm{N}$; volume averaged induced magnetic moment $\langle M\rangle$ from XMCD measurement in Fig. \ref{fig1}; interface magnetic moment $M_i$ \cite{Note1}; Py\textbar N interfacial exchange energy per interface atom $J_{ex}$ (Eq. \ref{jex_mcd}).}
\label{table1}
\end{table}

From the values of induced moments in Pd and Pt, we can make a step forward and estimate the interfacial exchange coupling energies for the two cases. 
Equating interatomic exchange energy $J_{ex}$ and Zeeman energy for an interface paramagnetic atom, we have (see Appendix \ref{paramagnets_section} for the derivation)

\begin{equation}
J_{ex}=\frac{1}{2} {\langle M\rangle\over \mu_B N_0 S}{t_\textrm{N}\over t_i} \label{jex_mcd}
\end{equation}

where $\langle M\rangle$ is the thickness-averaged paramagnetic moment, $N_0$ is the single-spin density of states (in eV$^{-1}$at$^{-1}$), S is the Stoner factor and $t_\textrm{i}=2*a/\sqrt{3}$ is the polarized interface-layer thickness \cite{Note1}. 
The $\nicefrac{1}{2}$ factor accounts for the fact that in XMCD measurements the N layer has both interfaces in contact with F. 
Under the simplifying assumption that all the magnetic moment is confined to the interface N layer and assuming experimental bulk susceptibility parameters for $\chi_v$, we obtain $J_{ex}^{Pd}=\textrm{42 meV}$ for Pd and $J_{ex}^{Pt}=\textrm{109 meV}$ for Pt (results and properties are summarized in Tab. \ref{table1}). 
Here the difference in estimated $J_{ex}$, despite roughly equal $M_{i}$, comes from the larger Stoner factor S for Pd.
A stronger interfacial exchange energy in Pt denotes a stronger orbital hybridization, yielding possibly a higher orbital character of the interfacial magnetic moment in the ferromagnetic Py counterpart \citep{Lee2007}.
For comparison, we consider the interatomic exchange parameters $J_{ex}$ in ferromagnetic Py and Co, investigated in Ref. \citep{Ghosh2012a}. 
$J_{ex}$ is estimated from the respective Curie temperatures $T_\textrm{C}$, through $J_{ex} \simeq 6 k_B T_C/\left(m/\mu_B\right)^2$, where $m$ is the atomic moment in \unit{$\mu_B$}/{at} (see Appendix \ref{ferromagnets_section}).
Experimental Curie temperatures of \unit[870]{K} and \unit[1388]{K} give $J_{ex}^{Co}=$\unit[293]{meV} for Co and $J_{ex}^{Py}=$\unit[393]{meV} for Py, which are of the same order of the value calculated for Pt (details about calculation in Appendix \ref{ferromagnets_section}). 

In the following, the effect of these static induced moments on the spin pumping damping of the heterostructures characterized will be discussed.

\subsection{FMR: damping enhancement}\label{FMR}

\begin{figure}
  \includegraphics[width=\columnwidth]{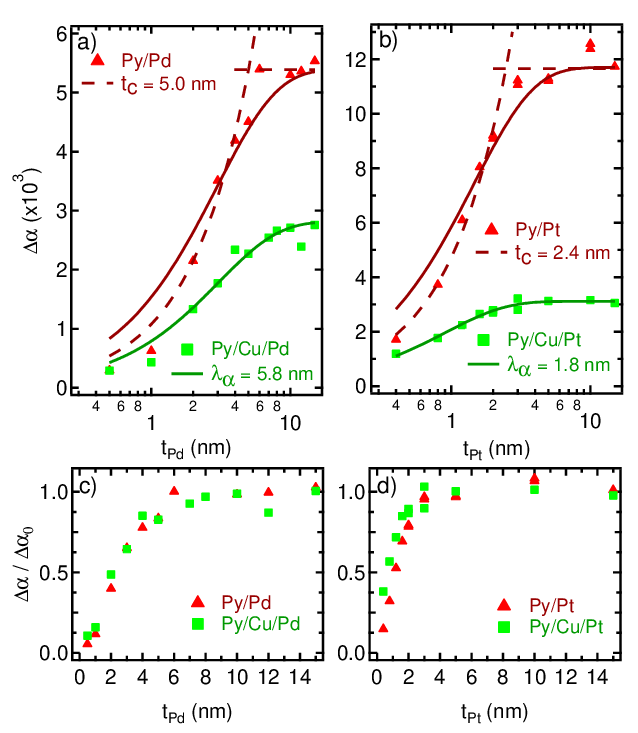}
  \caption{(Color online)
	Damping enhancement $\Delta\alpha$, due to pumped spin current absorption, as a function of thickness $t_{\textrm{N}}$ for Py(\unit[5]{nm})/N and Py(\unit[5]{nm})/Cu(\unit[3]{nm})/N heterostructures, with N = Pd($t_{\textrm{N}}$) (panels a,c), Pt($t_{\textrm{N}}$) (panels b,d). 
	Solid lines result from a fit with exponential function (Eq. \ref{expo}) with decay length $\lambda_\alpha$. 
	Dashed lines represents instead a linear-cutoff behavior (Eq. \ref{linear}) for $t_{\textrm{N}} < t_\textrm{c}$. 
	Please notice in panels a, c the x-axis is in logarithmic scale.
	In panels b, c the damping enhancement is normalized to the respective saturation value $\Delta\alpha_0$.
	}
	\label{fig2}
\end{figure}

The main result of our work is now shown in Figure \ref{fig2}.
In Fig. \ref{fig2} the damping enhancement $\Delta\alpha$ is plotted as a function of the spin-sink layer thickness $t_\textrm{N}$, for Py/Pd, Py/Cu/Pd (panels a, c) and Py/Pt, Py/Cu/Pt (panels b, d). 
The enhancement $\Delta\alpha$ is compared with the damping $\alpha$ of a reference structure Py(\unit[5]{nm})/Cu, excluding the sink layer N.
Each value of $\alpha$ results from established analysis of the linewidth of 11 FMR traces \citep{Ghosh2011,Ghosh2012a}, employing a \emph{g}-factor equal to 2.09 as a constant fit parameter for all samples. 

In Py/Cu/N systems (Fig. \ref{fig2}, green square markers), $\Delta\alpha$ rises with increasing $t_\textrm{N}$ thickness to similar saturation values $\Delta\alpha_0=0.0027$, 0.0031 for Pd and Pt, but reached on different length scales, given the different characteristic spin relaxation lengths of the two materials.
From the saturation value, an effective mixing conductance $g_\textrm{eff}^{\uparrow\downarrow}(\textrm{Py\textbar Cu/N})=\unit[7.2-8.3]{nm^{-2}}$ is deduced in the framework of standard spin pumping picture \citep{Tserkovnyak2002,Ghosh2011,Nan2015}, with Py saturation magnetization $\mu_0M_\textrm{s}=\unit[1.04]{T}$.
The fact that the spin-mixing conductance is not material dependent indicates that similar Cu\textbar N interfaces are formed.
The thickness dependence is well described by the \emph{exponential} function\citep{Foros2005,Ghosh2012a} 
\begin{equation}
\Delta\alpha(t_\textrm{N})=\Delta\alpha_{0}(1-\exp{(-2t_\textrm{N}/\lambda_{\alpha}^{N})}) \label{expo}
\end{equation}
as shown by the fit in Fig. \ref{fig2}a-b (continuous line).
As a result, exponential decay lengths $\lambda_{\alpha}^{Pt}=$ \unit[1.8]{nm} and $\lambda_{\alpha}^{Pd}=$ \unit[5.8]{nm} are obtained for Pt and Pd, respectively.

\begin{table}
\begin{tabular}{c|c|c|c|c} 
\textrm{N}	& $g_\textrm{eff}^{\uparrow\downarrow}(\textrm{Py\textbar Cu/N})$ & $\lambda_{\alpha}$ & $g_\textrm{eff}^{\uparrow\downarrow}(\textrm{Py\textbar N})$ & t$_c$ \\
&	[nm$^-2$] & [nm] & [nm$^-2$] & [nm] \\ \hline 
Pd & 7.2 & 5.8\footnotesize{$\pm$0.2} & 14 & 5.0\footnotesize{$\pm$0.3} \\ 
Pt & 8.3 & 2.4\footnotesize{$\pm$0.1} & 32 & 1.8\footnotesize{$\pm$0.2} \\ 
\end{tabular}
\centering
\caption{
Mixing conductance values extracted from the damping enhancement $\Delta\alpha$ at saturation in Fig. \ref{fig2}, and respective length scales (see text for details).}
\label{table2}
\end{table}

When the Pt, Pd spin-sink layers come into direct contact with the ferromagnetic Py, the damping enhancement $\Delta\alpha(t_\textrm{N})$ changes dramatically. 
In Py/N systems (Fig. \ref{fig2}a-b, triangle markers), the damping saturation values become $\Delta\alpha_0^{Pt}=0.0119$ and $\Delta\alpha_0^{Pd}=0.0054$ for Pt and Pd, respectively a factor $\sim$2 and $\sim$4 larger as compared to Py/Cu/N.
Within the spin-pumping description, a larger damping enhancement implies a larger spin-current density pumped out of the ferromagnet across the interface and depolarized in the sink.

\begin{figure}
  \includegraphics[width=\columnwidth]{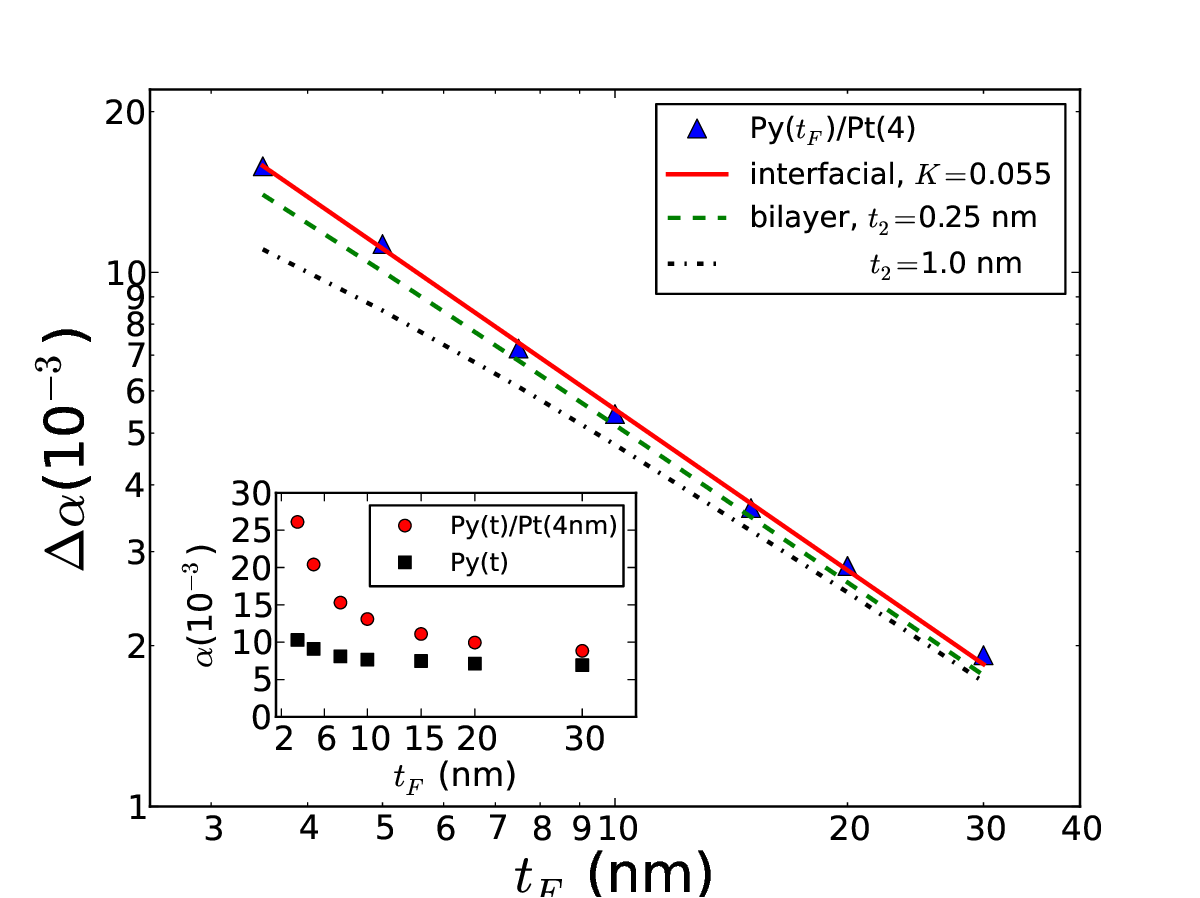}
  \caption{
	Logarithmic plot of the damping enhancement $\Delta\alpha$ (triangle markers) as a function of the Py layer thickness $t_\textrm{F}$, in Py(t$_\textrm{F}$)/Pt(\unit[4]{nm}). 
	Solid and dashed lines represents, respectively, fits according to the spin pumping (\emph{interfacial}) model $\Delta\alpha=K t_\textrm{F}^{-1}$ and to a $\alpha^{low}$(t$_\textrm{F}$)/$\alpha^{high}$($t_2$) \emph{bilayer} model, with $t_2=0.25, 1.0$\unit{nm}.
	\emph{Inset}: Gilbert damping $\alpha$ for Py(t$_\textrm{F}$) (square markers) and Py(t$_\textrm{F}$)/Pt(\unit[4]{nm}) (round markers). 
	}
	\label{figBiLayer}
\end{figure}

In Py/N heterostructures, because of the magnetic proximity effect, few atomic layers in N are ferromagnetically polarized, with a magnetic moment decaying with distance from the Py\textbar N interface.
The higher value of damping at saturation might therefore be interpreted as the result of a magnetic bi-layer structure, with a thin ferromagnetic N characterized by high damping $\alpha_{high}^{N}$ coupled to a low damping $\alpha_{low}^{F}$ ferromagnetic Py \citep{Song2004}.
To investigate whether damping is of bi-layer type, or truly interfacial, in Fig \ref{figBiLayer} we show the $t_\textrm{F}$ thickness dependence of the damping enhancement $\Delta\alpha$, for a Py($t_\textrm{F}$)/Pt(4 nm) series of samples.
The power law thickness dependence adheres very closely to $t_\textrm{F}^{-1}$, as shown in the logarithmic plot.
The assumption of composite damping for synchronous precession, as $\Delta\alpha(t_1)=(\alpha_1 t_1 + \alpha_2 t_2)/(t_1+t_2)$, shown here for $t_2=\textrm{\unit[0.25]{nm} and \unit[1.0]{nm}}$, cannot follow an inverse thickness dependence over the decade of $\Delta\alpha$ observed. 
Damping is therefore observed to have a pure interfacial character. 

In this case, the mixing conductances calculated from the saturation values are $g_\textrm{eff}^{\uparrow\downarrow}(\textrm{Py\textbar Pd})=\unit[14]{nm^{-2}}$ and $g_\textrm{eff}^{\uparrow\downarrow}(\textrm{Py\textbar Pt})=\unit[32]{nm^{-2}}$.
From \emph{ab initio} calculations within a standard spin-pumping formalism in diffusive films \citep{Zhang2011,Liu2014}, it is found $g_\textrm{eff}^{\uparrow\downarrow}(\textrm{Py\textbar Pd})=\unit[23]{nm^{-2}}$ for Pd and $g_\textrm{eff}^{\uparrow\downarrow}(\textrm{Py\textbar Pt})=\unit[22]{nm^{-2}}$ for Pt. 
Theoretical spin mixing conductance from a standard picture does reproduce the experimental order of magnitude, but it misses the 2.3 factor between the Py\textbar Pt and Py\textbar Pd interfaces.
Beyond a standard pumping picture, Liu and coworkers \citep{Liu2014} introduce spin-flipping scattering at the interface and calculate from first principles, for ideal interfaces in finite diffusive films: $g_\textrm{eff}^{\uparrow\downarrow}(\textrm{Py\textbar Pd})=\unit[15]{nm^{-2}}$, in excellent agreement with the experimental value here reported for Pd (Tab. \ref{table2}), and $g_\textrm{eff}^{\uparrow\downarrow}(\textrm{Py\textbar Pt})=\unit[25]{nm^{-2}}$.
Zhang \emph{et al.} \citep{Zhang2011} suggest an increase up to 25\% of the mixing conductance can be obtained by introducing magnetic layers on the Pt side.
The results here reported support the emerging idea that a generalized model of spin pumping including spin-orbit coupling and induced magnetic moments at F\textbar N interfaces may be required to describe the response of heterostructures involving heavy elements. 

\begin{figure}
	\includegraphics[width=\columnwidth]{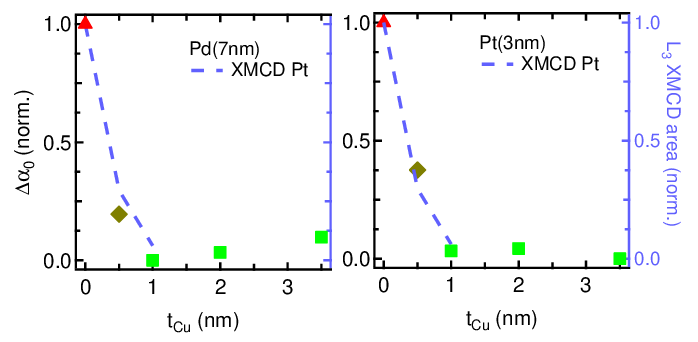}
	\caption{(Color online) Normalized damping enhancement $\Delta\alpha$ (left axis), due to spin pumping, as a function of interlayer thickness $t_{\textrm{Cu}}$ for Py(\unit[5]{nm})/Cu($t_{\textrm{Cu}}$)/N heterostructures, with N = Pd(\unit[7]{nm}), N = Pt(\unit[3]{nm}). The dashed line represents the XMCD signal (right axis) reported from inset in Fig. \ref{fig_XMCD_Cu}.}
	\label{fig_Alpha_Cu}
\end{figure}

The saturation value of damping enhancement at $\Delta\alpha_0$ as a function of the Cu interlayer thickness is shown in Fig. \ref{fig_Alpha_Cu} to follow the same trend of the XMCD signal (dashed line), reported from Fig. \ref{fig_XMCD_Cu}.
Indeed, it is found that the augmented $\Delta\alpha_0$ in Py/N junctions is drastically reduced by the insertion of \unit[0.5]{nm} Cu at the Py\textbar N interface \citep{Nan2015}, and the saturation of the Py/Cu/N configuration is already reached for \unit[1]{nm} of Cu interlayer.
As soon as a continuous interlayer is formed and no magnetic moment is induced in N, $\Delta\alpha_0$ is substantially constant with increasing Cu thickness. 

The N-thickness dependence of $\Delta\alpha(t_\textrm{N})$ in Py/N systems before saturation is addressed in the following.
At variance with the Py/Cu/N case, the thickness dependence of $\Delta\alpha$ is not anymore well described by an exponential behavior, as an exponential fit (with exponential decay length as only free fit parameter) fails to reproduce the increase of $\Delta\alpha$ towards saturation (solid lines in Fig. \ref{fig2}a-b). 
More rigorous fitting functions employed in spin pumping experiments, within standard spin transport theory \citep{Tserkovnyak2005,Boone2013,Boone2015}, cannot as well reproduce the experimental data (see Appendix. \ref{damping_section}).
It is worth mentioning that the same change of trend between the two configurations was observed for the same stacks with a \unit[10]{nm} thick Py layer (data shown in Appendix \ref{damping_section}, Fig. \ref{figS1}).
A change of the functional dependence of $\Delta\alpha$ on $t_\textrm{N}$ reflects a change in the spin-depolarization processes the pumped spin current undergoes, as for instance shown in Ref. \citep{Chen2015a} when interfacial spin-orbit coupling is introduced in the spin-pumping formalism.
Experimentally, a \emph{linear} thickness dependence with sharp cutoff has been shown to characterize spin-current absorption in spin-sink layers exhibiting ferromagnetic order at the interface, as reported for F$_1$/Cu/F$_2$($t_{\textrm{F}_2}$) junctions with F = Py, Co, CoFeB \citep{Ghosh2012a}. 
Given the presence of ferromagnetic order in N at the interface of F/N structures, the data are tentatively fit with a linear function 
\begin{equation}
\Delta\alpha = \Delta\alpha_{0}t_\textrm{N}/t_\textrm{c}^\textrm{N} \label{linear}
\end{equation}
This linear function better reproduce the sharp rise of $\Delta\alpha$ (dashed lines in Fig. \ref{fig2}a-b) and gives cutoff thicknesses $t_\textrm{c}^{Pt}=\unit[2.4\pm0.2]{nm}$ and $t_\textrm{c}^{Pd}=\unit[5.0\pm0.3]{nm}$ for Pt and Pd, respectively. 
The linearization is ascribed to the presence of ferromagnetic order in the paramagnetic Pd, Pt spin-sink layers at the interface with the ferromagnetic Py.
The linear trend extends beyond the thickness for which a continuous layer is already formed (less than \unit[1]{nm}), and, especially for Pd, far beyond the distance within the non-uniform, induced moment is confined (up to \unit[0.9]{nm}). 
In Ref. \citep{Ghosh2012a}, the cutoff $t_\textrm{c}$ in F/Cu/F heterostructures is proposed to be on the order of the transverse spin coherence length $\lambda_\textrm{J}$ in ferromagnetically ordered layers. 
$\lambda_\textrm{J}$ can be expressed in terms of the exchange splitting energy $J_{ex}$, 
\begin{equation}\label{EqLambdaJ}
\lambda_\textrm{J}= \frac{h v_g}{2J_{ex}}
\end{equation}
where $v_g$ is the electronic group velocity at the Fermi level.
This form, found from hot-electron Mott polarimetry \citep{Weber2001}, is expressed equivalently for free electrons as $\pi/|k^{\uparrow}-k^{\downarrow}|$, which is a scaling length for geometrical dephasing in spin momentum transfer \citep{Stiles2002}.
Electrons which enter the spin-sink at E$_F$ do so at a distribution of angles with respect to the interface normal, traverse a distribution of path lengths, and precess by different angles (from minority to majority or \emph{vice versa}) before being reflected back into the pumping ferromagnet.
For a constant $v_g$, it is therefore predicted that $t_\textrm{c}$ is inversely proportional to the exchange energy $J_{ex}$. 

\begin{figure}
	\includegraphics[width=0.8\columnwidth]{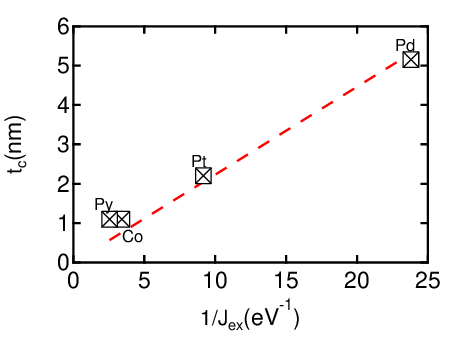}
	\caption{
	Effect of direct exchange strength on length scale of spin current absorption. 
	Cutoff thickness $t_\textrm{c}$ extracted from the $\Delta\alpha(t_\textrm{N})$ data in Fig. \ref{fig2} as a function of reciprocal interfacial exchange energy $1/J_{ex}$ extracted from XMCD in Fig. \ref{fig1}.
	Labels are given in terms of $J_{ex}$. 
	The Co and Py points are from Ref. \citep{Ghosh2012a}.}
	\label{fig3}
\end{figure}

In Figure \ref{fig3} we plot the dependence of the cutoff thickness $t_\textrm{c}^{\textrm{N}}$ upon the inverse of the estimated exchange energy $J_{ex}$ (Tab. \ref{table1}), as extracted from the XMCD measurements. 
A proportionality is roughly verified, as proposed for the transverse spin coherence length across spin polarized interfaces.
Under the simplistic assumption that $t_\textrm{c}=\lambda_\textrm{J}$, from the slope of the line we extract a Fermi velocity of $\sim\unit[0.1\cdot10^6]{m/s}$ (Eq. \ref{EqLambdaJ}), of the order of magnitude expected for the materials considered \citep{Dye1978,himpselAPL98}. 
These data show that, up to a certain extent, length scale for spin-current scattering shares common physical origin in ferromagnetic layers and paramagnetic heavy-metals, such as Pd and Pt, under the influence of magnetic proximity effect. 
This unexpected results is observed in spite of the fact that F$_1$/Cu/F$_2$ and F/N systems present fundamental differences.
In F/N structure, the induced moment in N is expected to be directly exchange coupled with the ferromagnetic counterpart.
Whereas in F$_1$/Cu/F$_2$, the magnetic moment in F$_2$ (off-resonance) are only weakly coupled with the precession occurring in F$_1$ (in-resonance), through spin-orbit torque and possible RKKY interaction. 
Magnetization dynamics in N might therefore be expected with its own pumped spin current, albeit, to the best of our knowledge, no experimental evidence of a dynamic response of proximity induced moments was reported so far.
From these considerations and the experimental findings, counter-intuitively the proximity-induced magnetic moments appear not to be involved in the production of spin current, but rather to contribute exclusively with an additional spin-depolarization mechanism at the interface.

\section{Conclusions}

We have investigated the effect of induced magnetic moments in heavy metals at Py/Pt and Py/Pd interfaces on the absorption of pumped spin currents, by analyzing ferromagnetic resonance spectra with varying Pt, Pd thicknesses. 
Static, proximity-induced magnetic moments amount to 0.32 and 0.3 $\mu_B/\textrm{atom}$ in Pd and Pt, respectively, at the interface with Py, as deduced from XMCD measurements taken at the L$_{2,3}$ edges.
We have shown that when the proximity induced moment in Pt and Pd is present, an onset of a linear-like thickness dependence of the damping is observed, in contrast with an exponential trend shown by Py/Cu/Pd and Py/Cu/Pt systems, for which no induced moment is measured. 
These results point to the presence of an additional spin-flip process occurring at the interface and to a change of the character of spin current absorption in the ultrathin Pd and Pt paramagnets because of the interfacial spin polarization.
The range of linear increase is proposed to be inversely proportional to the interfacial exchange energy in Py/Pt and Py/Pd, inferred from XMCD data. 

WEB acknowledges the Universit\'{e} Joseph Fourier and Fondation Nanosciences for his research stay at SPINTEC. This work was supported in part by the U.S. NSF-ECCS-0925829 and the EU EP7 NMP3-SL-2012-280879 CRONOS. MC is financed by Fondation Nanosciences.

\appendix

\section{N-thickness dependence}\label{damping_section}

\begin{figure}
  \includegraphics[width=0.9\columnwidth]{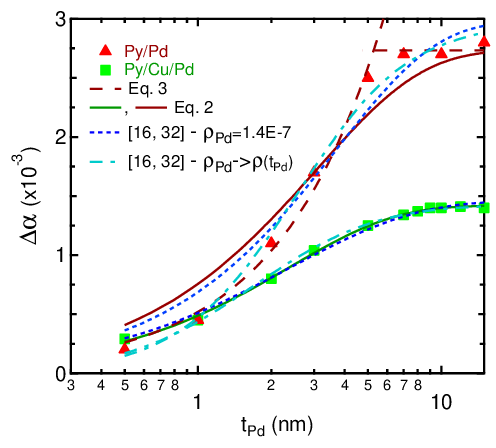}	
  \caption{
	Damping enhancement $\Delta\alpha$, due to pumped spin current absorption, as a function of thickness $t_{\textrm{Pd}}$ for Py(\unit[10]{nm})/Pd and Py(\unit[10]{nm})/Cu(\unit[3]{nm})/Pd heterostructures. 
	Solid lines result from a fit with exponential function (Eq. \ref{expo}) with decay $\lambda_\alpha$. 
	Dashed lines represents instead a linear-cutoff behavior (Eq. \ref{linear}) for $t_{\textrm{Pd}} < t_\textrm{c}$. 
	Short-dash and point-dash traces are fit to the data, employing equations from standard spin transport theory (see text for details) \citep{Boone2013,Boone2015}.
	In bottom panel, $\Delta\alpha$ is normalized to the respective saturation value.
	}
	\label{figS1}
\end{figure}

\begin{figure}
  \includegraphics[width=0.9\columnwidth]{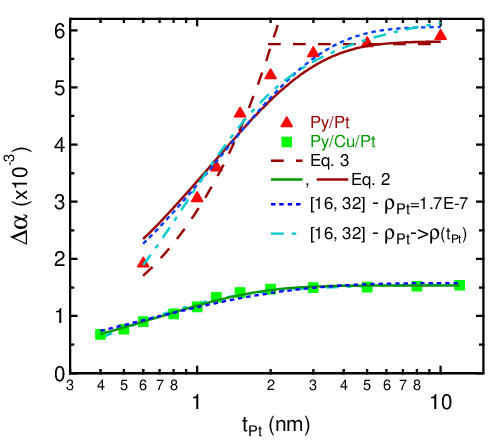}
  \caption{
	Damping enhancement $\Delta\alpha$, due to pumped spin current absorption, as a function of thickness $t_{\textrm{Pt}}$ for Py(\unit[10]{nm})/Pt and Py(\unit[10]{nm})/Cu(\unit[3]{nm})/Pt heterostructures. 
	Solid lines result from a fit with exponential function (Eq. \ref{expo}) with decay $\lambda_\alpha$. 
	Dashed lines represents instead a linear-cutoff behavior (Eq. \ref{linear}) for $t_{\textrm{Pt}} < t_\textrm{c}$. 
	Short-dash and point-dash traces are fit to the data, employing equations from standard spin transport theory (see text for details) \citep{Boone2013,Boone2015}.
	In bottom panel, $\Delta\alpha$ is normalized to the respective saturation value.
	}
	\label{figS2}
\end{figure}

In order to confirm the results presented in the manuscript, additional sample series with thicker Py layer were fabricated and measured.
The experimental results for \unit[10]{nm} thick Py layer are shown in Fig.s \ref{figS1} and \ref{figS2} for Pd and Pt, respectively.
We have presented the data here, rather than including them with the other plots in Figure \ref{fig2}, to keep the figures from being overcrowded.
As expected when doubling the ferromagnet thickness, the saturation values $\Delta\alpha_0$ are about half of those measured for \unit[5]{nm} Py (Fig. \ref{fig2}).
Confirming the data presented in the manuscript, it is observed again a change of thickness dependence of $\Delta\alpha(t_{\textrm{N}})$, from \emph{exponential} for Py/Cu/N (solid lines; Eq. \ref{expo}, $\lambda_{\alpha}=$ \unit[4.8]{nm} and \unit[1.4]{nm} for Pd and Pt respectively) to \emph{linear}-like for Py/N (dashed lines; Eq. \ref{linear}, t$_c=$ \unit[5.3]{nm} and \unit[2]{nm} for Pd and Pt respectively). 

The experimental data are also fitted with a set of equations derived from standard theory of diffusive spin transport \citep{Tserkovnyak2005,Boone2013,Boone2015}, describing the the dependence of $\Delta\alpha$ on the thickness of adjacent metallic layers (either N or Cu/N in our case) as follow 
\begin{equation}
\Delta\alpha = \frac{\gamma \hbar}{4\pi M_\textrm{s} t_{FM}} \frac{g^{\uparrow\downarrow}}{1+\nicefrac{g^{\uparrow\downarrow}}{g_{ext}^x}}
\end{equation}
with (Eq. 7 in Ref. \citep{Boone2015}, and Eq. 6 in Ref. \citep{Boone2013})
\begin{equation}
\begin{aligned}
g_{ext}^{N} &= g_{N}\tanh{\nicefrac{t_N}{\lambda_{sd}^{N}}}\\
g_{ext}^{Cu/N} &= g_{Cu} \frac{g_{Cu}\coth{\nicefrac{t_N}{\lambda_{sd}^{N}}} + g_{N}\coth{\nicefrac{t_{Cu}}{\lambda_{sd}^{Cu}}}}{g_{Cu}\coth{\nicefrac{t_N}{\lambda_{sd}^{N}}}\coth{\nicefrac{t_{Cu}}{\lambda_{sd}^{Cu}}} + g_{N}}
\end{aligned}
\end{equation}
where $g_x=\nicefrac{\sigma_x}{\lambda_{sd}^{x}}$, $\sigma_x$ and $\lambda_{sd}^{x}$ are the electrical conductivity and spin diffusion length of the non magnetic layer \emph{x}. 
For the thin Cu layer, we used a resistivity $\rho_\textrm{Cu}=\unit[1\times 10^{7}]{\Omega m}$ and a spin diffusion length $\lambda^\textrm{Cu}_{sd}=\unit[170]{nm}$ \citep{Boone2013}. 
For the Pt and Pd layers, two fitting models in which the conductivity of the films is either constant or thickness dependent are considered, as recently proposed by Boone and coworkers \citep{Boone2015}. 
The values of conductivity, as taken directly from Ref. \citep{Boone2015}, will influence the spin diffusion length $\lambda^{N}_{sd}$ and spin mixing conductance $g^{\uparrow\downarrow}$ resulting from the fit, but will not affect the conclusions drawn about the overall trend. 
When a constant resistivity is used (short-dash, blue lines), the model basically corresponds to the simple exponential function in Eq. \ref{expo}. 
It nicely reproduces the data in the indirect contact case (Py/Cu/N) for both Pd (Fig. \ref{figS1}) and Pt (Fig. \ref{figS2}), but it fails to fit the direct contact (Py/N) configuration. 
When a thickness dependent resistivity of the form $\rho_\textrm{N}= \rho_\textrm{N}^{b} + \nicefrac{\rho_\textrm{N}^{s}}{t_\textrm{N}}$ is used (dash-point, cyan lines)  \citep{Boone2015}, in Py/Cu/N systems, no significant difference with the other functions is observed for Pt, while for Pd a deviation from experimental trend is observed below \unit[1.5]{nm}. 
In Py/N systems, the fit better describes the rise at thicknesses shorter than the characteristic relaxation length, while
deviates from the data around the saturation range. 

Models from standard spin transport theory cannot satisfactorily describe the experimental data for the direct contact Py/N systems. 
For this reason a different mechanism for the spin depolarization processes has been proposed, considering the presence of induced magnetic moments in N in contact with the ferromagnetic layer.

\section{Interfacial interatomic exchange}
 
\subsection{Paramagnets}\label{paramagnets_section}

We will show estimates for exchange energy based on XMCD-measured moments in [Py/(Pt, Pd)]$_{repeat}$ superlattices.  Calculations of susceptibility are validated against experimental data for Pd and Pt.  Bulk susceptibilities will be used to infer interfacial exchange parameters $J^i_{ex}$.

\paragraph{Pauli susceptibility}

For an itinerant electron system characterized by a density of states at the Fermi energy $N_0$, if an energy $\Delta E$ splits the spin-up and spin-down electrons, the magnetization resulting from the (single-spin) exchange energy $\Delta E$ is

\be M = \mu_B\left(N^\uparrow-N^\downarrow\right)=2\mu_B N_0 S \Delta E \ee

where $N_0$ is the density of states in $\#/\textrm{eV/at}$, $S$ is the Stoner parameter, and $2\Delta E$ is the exchange {\it splitting} in eV.  Moments are then given in $\mu_B/\textrm{at}$.  Solving for $\Delta E$,

\be  \Delta E = {M\over 2\mu_B N_0 S} \ee

If the exchange splitting is generated through the application of a magnetic field, $\Delta E=\mu_B H$,

\be  \mu_B H = {M\over 2\mu_B N_0 S} \ee

and the dimensionless volume magnetic susceptibility can be expressed

\be \chi_v \equiv {M\over H} =  2\mu_B^2 N_0\:S \label{chi_eq}\ee

In this expression, the prefactor can be evaluated through

\be \mu_B^2=\textrm{59.218 eV\AA}^3\ee  

so with $N_0[=]$/eV/at, $\chi_v$ takes units of volume per atom, and is then also called an atomic susceptibility, in cm$^3$/at, as printed in Ref \citep{l-b-paramag-4d5d}.

\paragraph{Molar susceptiblity}  Experimental values are tabulated as molar susceptibilities.  The atomic susceptibility $\chi_v$ can be contrasted with the mass susceptibility $\chi_m$ and molar susceptibility $\chi_{mol}$

\be \chi_{mass} = {\chi_v\over\rho}\qquad \chi_{mol}={\textrm{ATWT}\over\rho}\chi_v \ee

where ATWT is the atomic weight (g/mol) and $\rho$ is the density (g/cm$^3$).  These have units of $\chi_{mass}[=]\textrm{cm}^3/\textrm{g}$ and $\chi_{mol}[=]\textrm{cm}^3/\textrm{mol}$.  The molar susceptibility $\chi_{mol}$ is then

\be \chi_{mol} = 2\mu_B^2 N_0 N_A\:S\ee

in cm$^3$/mol, where $\mu_B$ is the Bohr magneton, and  

\be 2 N_0 S = {\chi_{mol}\over N_A\mu_B^2}\label{expt_val}\ee  

Eq. \ref{expt_val} provides a convenent method to estimate experimental unknowns, the density of states $N_0$ and Stoner parameter $S$, from measurements of $\chi_{mol}$.  

{\it Example:} for Pd, the low-temperature measurement (different from the room-temperature measurement in Table \ref{table1}) is $\chi_{mol}\sim\unit[7.0\times\textrm{10}^{-4}]{cm^3/mol}$.  In the denominator, $(N_A\mu_B^2)=\textrm{2.622}\times\textrm{10}^{-6}\textrm{Ry}\cdot\textrm{cm}^3/\textrm{mol}$,  The value $2 N_0 S$ consistent with the experiment is 266/(Ry-at) or 19.6/(eV-at).  For the tabulated measurement of $S=\textrm{9.3}$, the inferred density of states is then $N_0=\textrm{1.05/eV/at}$.

\paragraph{Interfacial exchange}

We can assume that the Zeeman energy per interface atom is equal to its exchange energy, through the Heisenberg form

\be {M_p^2\over\chi_v}V_{at} = 2 J_{ex}^i s_f s_p \label{energy_balance}\ee

where $M_p$ is the magnetization of the paramagnet, with the atomic moment of the paramagnet $m_p$ in terms of its per-atom spin $s_p$,

\be M_p = {m_p\over V_{at}}\qquad m_p = 2\mu_B s_p \label{mps}\ee 

$V_{at}$ is the volume of the paramagnetic site, $s_{f,p}$ are the per-atom spin numbers for the ferromagnetic and paramagnetic sites, and $J_{ex}^i$ is the (interatomic) exchange energy acting on the paramagnetic site from the ferromagnetic layers on the other side of the interface.  Interatomic exchange energy has been distinguished from intraatomic (Stoner) exchange involved in flipping the spin of a single electron.  Rewriting Eq \ref{energy_balance},

\be {M_p^2\over\chi_v}V_{at} = 2 J_{ex}^i s_f {M_p\over 2\mu_B}V_{at} \ee

if $s_f=1/2$, appropriate for $4\pi M_s\sim\textrm{10 kG}$,

\be J_{ex}^i =2 \mu_B{M_p\over\chi_v} \ee

and substituting for $\chi_v$ through Eq \ref{chi_eq},

\be J_{ex}^i = {M_p\over \mu_B N_0 S} \ee

In the XMCD experiment, we measure the thickness-averaged magnetization as $<M>$ in a [$F/N$]$_{n}$ superlattice.  We make a simplifying assumption that the exchange acts only on nearest-neighbors and so only the near-interface atomic layer has a substantial magnetization.  We can then estimate $M_p$ from $<M>$ through

\be <M> t_p = 2 M_p t_i \ee

where $t_i$ is the polarized interface-layer thickness of $N$ \cite{Note1}. 
Since the interface exists on both sides of the $N$ layer, $2 t_i$ is the thickness in contact with $F$. 
Finally,

\be J_{ex}={1\over 2}{<M>\over \mu_B N_0 S}{t_p\over t_i} \ee

The exchange energy acting on each interface atom, from all neighbors, is $J_{ex}^{Pt}=\textrm{109 meV}$ for Pt and $J_{ex}^{Pd}=\textrm{42 meV}$ for Pd. 
Per nearest neighbor for an ideal F/N(111) interface, it is $J^{\textrm{Py\textbar Pt}}=\textrm{36 meV}$ and $J^{\textrm{Py\textbar Pd}}=\textrm{14 meV}$. 
Per nearest neighbor for an intermixed interface (6 nn), the values drop to 18 meV and 7 meV, respectively. 

Since explicit calculations for these systems are not in the literature, we can compare indirectly with theoretical values. 
Dennler\citep{dennlerPRB05} showed that at a ($3d$)F/($4d$)N interface (e.g. Co/Rh), there is a geometrical enhancement in the moment induced in $N$ per nearest-neighbor of $F$. 
The $4d$ $N$ atoms near the $F$ interface have larger induced magnetic moments per nn of $F$ by a factor of four. 
Specific calculations exist of $J^{\textrm{F\textbar N}}$ (per neighbor) for dilute Co impurities in Pt and dilute Fe impurties in Pd \citep{ebertPRB10}. 
$J^{Fe-Pd}\sim\textrm{3 meV}$ is calculated, roughly independent of composition up to 20\% Fe. 
If this value is scaled up by a factor of four, to be consistent with the interface geometry in the XMCD experiment, it is $\sim$ 12 meV, comparable with the value for Pd, assuming intermixing. 
Therefore the values calculated have the correct order of magnitude.

\subsection{Ferromagnets}\label{ferromagnets_section}

The Weiss molecular field,

\be H_W = \beta M_s \ee

where $\beta$ is a constant of order 10$^{3}$, can be used to give an estimate of the Curie temperature, as 

\be T_C = {\mu_B g_J J\left(J+1\right)\over 3 k_B} H_W\ee

Density functional theory calculations have been used to estimate the molecular field recently\citep{brunoPRB01,ebertPRB10}; for spin type, the $J\left(J+1\right)$ term is substutited with $<s>^2$, giving an estimate of

\be T_C= {2<s>^2 J_0\over 3 k_B}\ee

where $<s>$ is the number of spins on the atom as in Eq \ref{mps}; see the text by St\"{o}hr and Siegmann\citep{ss1a}. 
$<s>$ can be estimated from $m=$1.07$\mu_B$ for Py and 1.7$\mu_B$ for Co, respectively. 
Then

\be J_0 \simeq {6 k_B T_C\over \left(m/\mu_B\right)^2} \label{J0fm}\ee

with experimental Curie temperatures of 870 and 1388 K, respectively, gives estimates of $J_0=\textrm{293 meV}$ for Co and $J_0=\textrm{393 meV}$ for Py.

Note that there is also a much older, simpler method. 
Kikuchi\citep{kikuchiAOP58} has related the exchange energies to the Curie temperature for FCC lattices through

\be J=\textrm{0.247}k_B T_C\ee

Taking 12 NN, $\textrm{12}J$ gives a total energy of 222 meV for Py (870 K) and 358 meV for FCC Co (1400K), not too far off from the DFT estimates.

\paragraph{Other estimates}

The $J_0$ exchange parameter is interatomic, describing the interaction between spin-clusters located on atoms. 
Reversing the spin of one of these clusters would change the energy $J_0$. 
The Stoner exchange $\Delta$ is different, since it is the energy involved in reversing the spin of a single electron in the electron sea. 
Generally $\Delta$ is understood to be greater than $J_0$ because it involves more coloumb repulsion; interatomic exchange can be screened more easily by $sp$ electrons.  

This exchange energy is that which is measured by photoemission and inverse photoemission. 
Measurements are quite different for Py and Co.  Himpsel\citep{himpselAPL98} finds an exchange splitting of $\Delta=\textrm{270 meV}$ for Py, which is not too far away from the Weiss $J_0$ value.  For Co, however, the value is between 0.9 and 1.2 eV, different by a factor of four. 
For Co the splitting needs to be estimated by a combination of photoemission and inverse photoemission because the splitting straddles $E_F$.\citep{schneiderPRL90}.

For comparison with the paramagnetic values of $J_{ex}^i$, we use the $J_0$ estimates, since they both involve a balance between Zeeman energy (here in the Weiss field) and Heisenberg interatomic exchange. 
Nevertheless the exchange splitting $\Delta_{ex}$ is more relevant for the estimate of $\lambda_c = h v_g/(2\Delta_{ex})$. 
For Py, the predicted value of $\lambda_c$ from the photoemission value (through $\lambda_c = \pi/|k^\uparrow-k^\downarrow|$) is 1.9 nm, not far from the experimental value of 1.2 nm. 

\bibliographystyle{apsrev}
\bibliography{References}

\end{document}